\documentclass[runningheads]{llncs}
\usepackage[T1]{fontenc}
\usepackage{graphicx}
\usepackage{amsmath,amssymb,amsfonts}
\usepackage{algorithmic}
\usepackage{graphicx}
\usepackage{textcomp}
\usepackage{xcolor}
\usepackage{tikz}
\usepackage{xspace}
\usepackage{graphics}
\usepackage{subfigure}
\usepackage{multirow}
\usepackage{marvosym}
\usepackage{pifont}
\usepackage[ruled,vlined,linesnumbered]{algorithm2e}
\usepackage{pifont}
\usepackage{titlecaps}
\usepackage[T1]{fontenc}
\usepackage{colortbl}
\usepackage{graphicx,xcolor}
\usepackage{listings}
\usepackage{enumitem}
\usepackage[numbers]{natbib}

\newcommand{\xstore}{SynchroStore\xspace}

\begin{document}
\title{SynchroStore: A Cost-Based Fine-Grained Incremental Compaction for Hybrid Workloads}
%
%
\author{Yinan Zhang\inst{1} \and
Huiqi Hu\inst{2} \and
Xuan Zhou\inst{3}}
\authorrunning{F. Author et al.}
%
\institute{East China Normal University \email{ynzhang@stu.ecnu.edu.cn}
\and
East China Normal University \email{hqhu@dase.ecnu.edu.cn} \\
\and
East China Normal University \email{xzhou@dase.ecnu.edu.cn}
}





%
\maketitle              
\begin{abstract}

This study proposes a novel storage engine, \xstore, designed to address the inefficiency of update operations in columnar storage systems based on Log-Structured Merge Trees (LSM-Trees) under hybrid workload scenarios. While columnar storage formats demonstrate significant query performance advantages when handling large-scale datasets, traditional columnar storage systems face challenges such as high update complexity and poor real-time performance in data-intensive applications. \xstore introduces an incremental row storage mechanism and a fine-grained row-to-column transformation and compaction strategy, effectively balancing data update efficiency and query performance. The storage system employs an in-memory row storage structure to support efficient update operations, and the data is converted to a columnar format after freezing to support high-performance read operations.

The core innovations of \xstore are reflected in the following aspects: (1) the organic combination of incremental row storage and columnar storage; (2) a fine-grained row-to-column transformation and compaction mechanism; (3) a cost-based scheduling strategy. These innovative features allow \xstore to leverage background computational resources for row-to-column transformation and compaction operations, while ensuring query performance is unaffected, thus effectively solving the update performance bottleneck of columnar storage under hybrid workloads. Experimental evaluation results show that, compared to existing columnar storage systems like DuckDB, \xstore exhibits significant advantages in update performance under hybrid workloads.

\keywords{LSM-tree  \and Compaction \and Column Store.}
\end{abstract}

\section{Background}\label{dasexstore:problem}

With the exponential growth of data scale and the increasing complexity of computational demands, modern storage system architectures have exhibited significant diversification trends to meet the specific needs of different application scenarios. In traditional row-store systems, data is stored row by row, which demonstrates high efficiency in handling frequent data insertion and update operations. However, with the advent of the big data era and the surge in complex query demands, columnar storage systems~\cite{farber2012sap, stonebraker2018c, schulze2024clickhouse} have gradually emerged as superior solutions, particularly excelling in scenarios such as large-scale data scanning, aggregation computations, and analytical processing.

\subsection{Mixed Workload Challenges in Columnar Storage Systems}

Columnar storage systems organize data by columns, a design that provides significant advantages in specific query patterns, especially in data warehousing and online analytical processing (OLAP) systems. Compared to row-store systems, columnar storage systems can significantly improve query performance, primarily because they only need to read the columns involved in the query rather than loading entire rows~\cite{abadi2008column}. This design enables columnar storage systems to exhibit notable performance advantages when handling large-scale datasets, particularly in executing complex aggregation, filtering, and computational operations. Additionally, since data within the same column shares the same data type characteristics, columnar storage systems can achieve more efficient data compression, which plays a critical role in storage optimization and query performance enhancement~\cite{abadi2013design}.

However, despite the significant advantages of columnar storage in query performance and data compression, it faces severe challenges in handling data updates and write operations. Traditional row-store systems have inherent advantages in data updates because their row-based storage allows update operations to modify only the relevant rows~\cite{ramakrishnan2003database}. In contrast, update operations in columnar storage systems are more complex: when a column's data changes, the system often needs to reorganize the entire column, which not only increases the complexity of update operations but may also impact real-time system performance~\cite{idreos2007updating}.

In practical application scenarios, user requests often exhibit high complexity, typically comprising a mix of query and update requests. Particularly in real-time data processing fields such as the Internet of Things (IoT), financial transactions, social networks, and e-commerce platforms, systems need to respond quickly to dynamic data changes~\cite{kemper2011hyper}. In these systems, data continuously undergoes updates, modifications, deletions, and insertions, where the efficiency and timeliness of update operations are crucial for ensuring system stability, maintaining data consistency, and supporting business decisions. For example, financial trading systems require real-time updates to transaction records; social platforms need to reflect user activities promptly; and e-commerce platforms must track inventory and order statuses in real time. These demands require database systems to complete data updates in extremely short timeframes while ensuring the real-time accuracy of query results~\cite{ozsu1999principles}. If columnar storage systems cannot effectively support data update operations, they may fail to reflect the latest data states in time, potentially leading to data inconsistencies or decision-making errors, ultimately affecting user experience and business operations.

Modern database systems often face mixed workloads, comprising both complex data queries and analytical operations (OLAP) as well as frequent data updates and transaction processing (OLTP). Particularly in big data application scenarios, database systems need to handle large volumes of query and update requests simultaneously~\cite{chaudhuri1997overview}. Therefore, effectively supporting mixed workloads has become a key challenge in the design of columnar storage systems~\cite{plattner2009common}.

Traditional columnar storage systems are primarily optimized for query performance, while their capabilities in data updates are notably lacking. This design flaw makes it difficult for columnar storage systems to effectively address the demands of mixed workloads. When columnar storage systems cannot process update operations promptly, they exhibit significant performance bottlenecks in scenarios requiring rapid reflection of data changes. Thus, to support large-scale data analysis while achieving efficient real-time update processing, columnar storage systems must incorporate efficient update mechanisms, which are crucial for supporting mixed workloads~\cite{larson2011sql}.

To address this challenge, columnar storage systems primarily adopt two update strategies: in-place updates and incremental updates. In-place updates rewrite existing data to achieve updates on the fly, but this method incurs substantial update overhead and often lacks flexibility in scenarios with high real-time update demands. Therefore, most modern columnar storage engines tend to adopt incremental update strategies. Incremental updates mark old rows for deletion and insert new rows, avoiding the overhead of rewriting entire columns with each update. Currently, incremental updates are implemented in two main ways: incremental row updates and incremental columnar updates~\cite{lamb2012vertica}.

However, the adoption of incremental update strategies inevitably introduces new problems: after updates, incremental data must be merged with the original data to obtain a complete data view, a process that negatively impacts read performance. To minimize the impact of incremental updates on read performance, some columnar storage engines (e.g., ClickHouse~\cite{schulze2024clickhouse}, DuckDB~\cite{raasveldt2019duckdb}) choose to store incremental data in columnar format. For instance, in ClickHouse, when a column update is executed, the system inserts a delete marker to flag the old row and inserts the updated new row, which is stored in columnar format. However, this approach requires converting incremental data into columnar format during updates, which adversely affects update efficiency. Additionally, in single-row granularity update modes, this approach tends to generate data fragmentation, resulting in small, non-contiguous columnar files that cannot fully leverage the advantages of columnar storage systems in column access locality. Furthermore, periodically merging these small columnar files into larger ones introduces additional system overhead. Given the inherent difficulties of columnar formats in updates, adopting incremental columnar update strategies in mixed workload scenarios may severely constrain system update performance.

To efficiently support update operations, some columnar engines adopt incremental row storage for database updates (e.g., Kudu~\cite{lipcon2015kudu}, Doris~\cite{Doris}, ADB~\cite{adb}, TiFlash~\cite{tidb}). In this method, newly added row data is first written into an incremental row storage structure and then batch-converted into columnar format. Although this approach improves update speed and reduces data fragmentation, it inevitably impacts read performance. This is because traditional columnar engines cannot directly execute queries on row-stored data; they must convert incremental row data into columnar data before or during query execution, thereby increasing query overhead. Additionally, row-stored data cannot leverage SIMD technology for query acceleration, and its query speed is inherently slower than columnar data. However, the experimental results of this study show that if the proportion of incremental row-stored data can be controlled to remain at a low level, the impact of incremental row storage on the read performance of columnar storage systems can be effectively mitigated. The following sections will further analyze this phenomenon through experiments.

\subsection{Problem Analysis}

This study experimentally observes the impact of incremental columnar and incremental row storage on query performance. To this end, the study implements both incremental columnar and incremental row update methods in the storage system \xstore and compares their performance. In the experiment, 1.5 million rows of data are imported into \xstore, and incremental updates are performed at different ratios. After the updates, a query operation (Select col1 from table1) is executed on \xstore, and both update latency and query latency are measured. The experimental results are shown in Figure~\ref{dasexstore:problem_update}.

\begin{figure}[t]
	\center		
	\hspace{-6mm}
	\subfigure[Update Overhead]{
		\label{dasexstore:cost_update}
		\raisebox{0mm}{\includegraphics[width=0.45\linewidth]{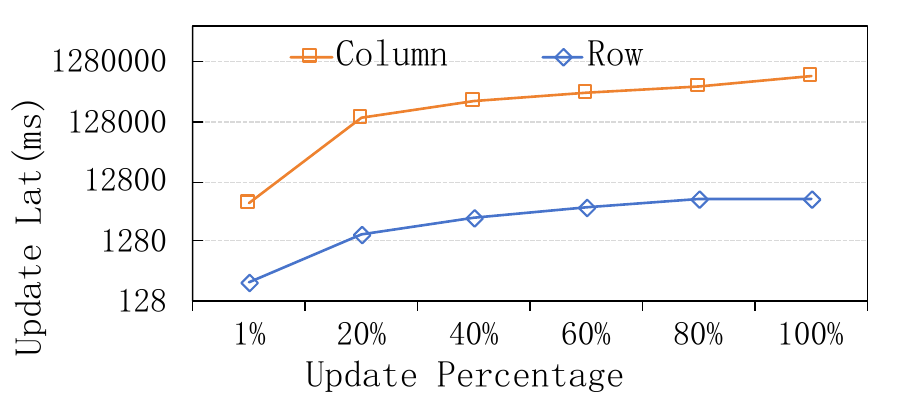}}
	}
	\subfigure[Impact of Updates on Performance]{
		\label{dasexstore:impact_update}
		\raisebox{0mm}{\includegraphics[width=0.45\linewidth]{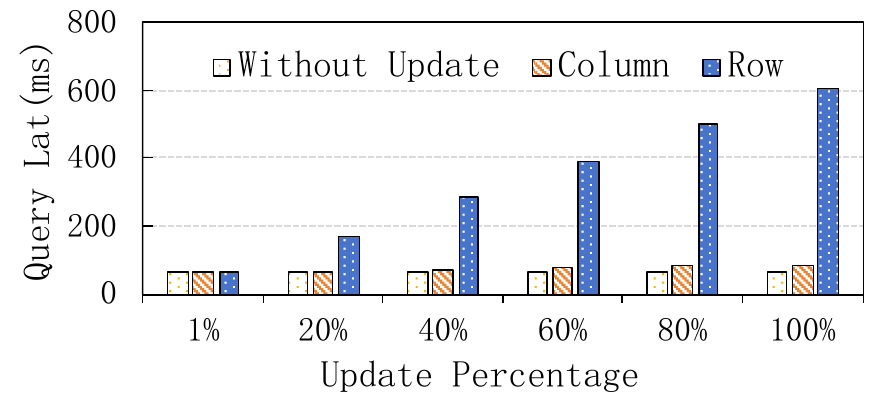}}
	}
	\caption{Incremental Update Overhead and Performance Impact in Columnar Engines}
    \label{dasexstore:problem_update}
\end{figure}

Figure~\ref{dasexstore:cost_update} shows the latency of incremental updates at different update ratios. The x-axis represents the update ratio (e.g., 60\% means randomly selecting 60\% of the data for incremental updates), and the y-axis represents update latency. The experimental results indicate that incremental row updates are significantly more efficient than incremental columnar updates. This phenomenon is primarily due to the fact that, during incremental columnar updates, each newly inserted row requires decompressing, updating, and re-encoding the incremental columnar structure, which incurs substantial overhead. In contrast, incremental row updates, due to their data structure characteristics, can complete update operations more efficiently.

Figure~\ref{dasexstore:impact_update} shows the changes in query latency after updates. The x-axis represents the data update ratio, and the y-axis represents read latency. The experimental results reveal that as the update ratio increases, the read latency of the \xstore system using incremental row storage rises significantly, while the read performance of the unupdated system remains relatively stable. In contrast, the read latency of the \xstore system using incremental columnar storage increases less dramatically. This difference is mainly due to the inherent query-friendly nature of incremental columnar formats, whereas incremental row updates require converting row-stored data into columnar format during query execution, which incurs substantial overhead. However, the study also finds that when the proportion of incremental row-stored data is low relative to the total data volume, its impact on overall query performance is relatively limited.

\subsection{Challenges}

The experimental results indicate that incremental row storage mechanisms significantly impact the read performance of columnar storage systems. To effectively mitigate the negative effects of incremental row storage on system read performance, this study proposes a background-thread-based periodic conversion mechanism. By periodically converting incremental row-stored data into columnar format, the scale of incremental row-stored data can be controlled, maintaining efficient update performance while leveraging the advantages of both incremental row storage for updates and columnar storage for reads. However, this approach faces the following challenges in practical applications:

\begin{enumerate}
    \item \textbf{Resource Consumption and Performance Fluctuations Due to Row-to-Column Conversion:}
    Row-to-column conversion is a resource-intensive operation, especially when converting large volumes of incremental row-stored data into columnar format at once. This process consumes significant computational resources and may compete with query operations for system resources, leading to performance fluctuations. Under high load conditions, row-to-column conversion operations may increase query latency, making the system unstable during query execution. Particularly when the system is handling high-concurrency queries, row-to-column conversion may cause query response times to rise significantly. Therefore, balancing resource allocation between row-to-column conversion operations and query requests to ensure stable performance under high load is a critical challenge.

    \item \textbf{High Overhead of Compaction Between Incremental Data and Baseline Data:}
    Simply converting incremental row-stored data into columnar format does not fully eliminate invalid data caused by update operations. Therefore, periodic compaction of incremental data with baseline data (Baseline data) is necessary. Since update operations are typically random, there is a high overlap between incremental data and baseline data, which significantly increases the overhead of the compaction process and further burdens the system. Additionally, the compaction process itself consumes substantial computational resources, especially when the scale of incremental data is large, potentially causing the system to be unresponsive to normal query requests for extended periods. This not only increases storage overhead but also significantly reduces system throughput and response speed.

    \item \textbf{Impact of Client Load Fluctuations on Row-to-Column Conversion and Compaction Operations:}
    Client load typically exhibits dynamic fluctuations, especially in scenarios with large-scale concurrent queries and operations. When the system is under high load, executing row-to-column conversion or data compaction operations may exacerbate performance degradation. For example, during peak query request periods, if the system simultaneously performs large-scale row-to-column conversion or data compaction operations, resource contention may lead to significant increases in query latency or even cause system hangs or timeouts. Therefore, when scheduling row-to-column conversion and compaction operations in background threads, the fluctuation characteristics of client load must be carefully considered to ensure these operations are primarily executed during low-load periods, minimizing their impact on high-load periods. This requires designing more intelligent scheduling mechanisms for row-to-column conversion and compaction operations to ensure their smooth execution while maintaining query performance.
    
\end{enumerate}

To address the aforementioned problems and challenges in the update process of columnar storage engines, this study designs and develops a new columnar storage engine, \xstore. By integrating the technical advantages of incremental row storage and incremental columnar storage, \xstore achieves optimized adaptation to mixed workloads.

\section{Overview of \xstore Design}

\subsection{Solution Approach}

\begin{figure}[h]
    \centering
    \includegraphics[width=0.75\linewidth]{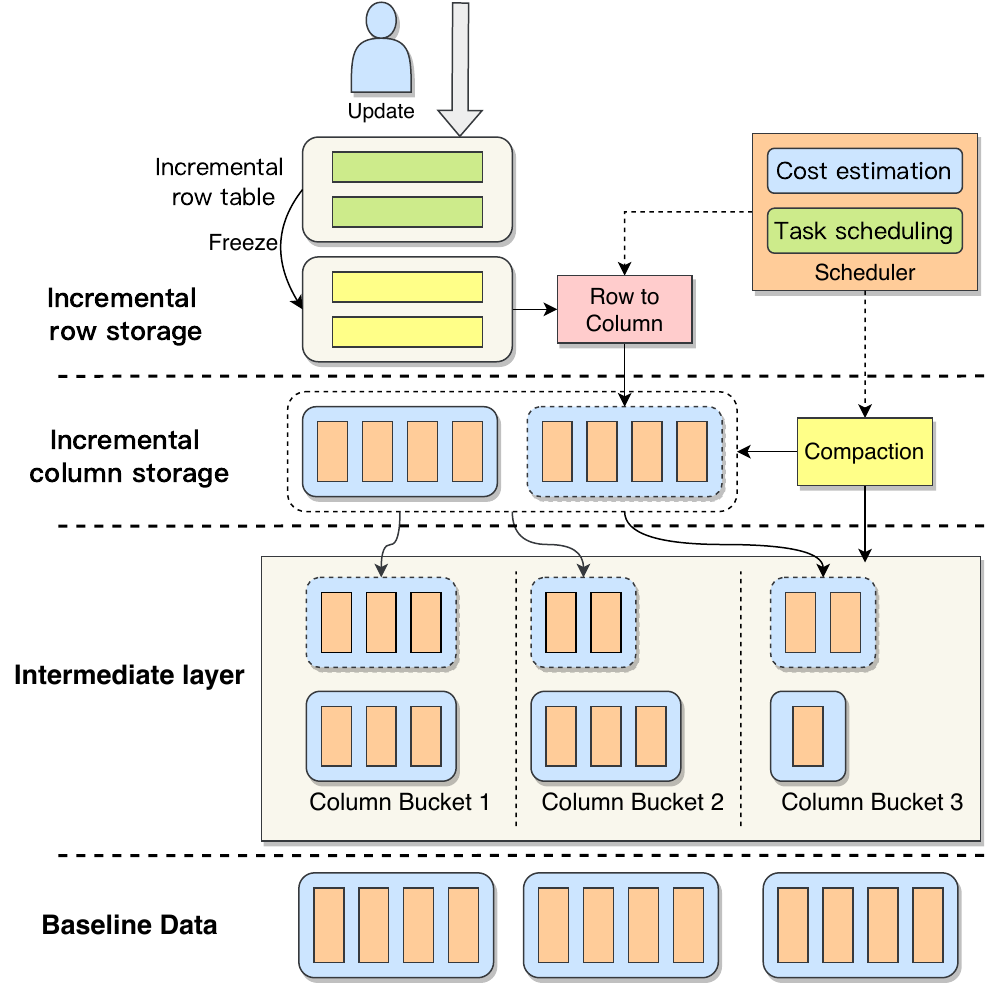}
    \caption{Overview of \xstore Design}
    \label{dasexstore:overview}
\end{figure}

The design of this study is based on the Log-Structured Merge-Tree (LSM-Tree). In an LSM-Tree, only one active in-memory table supports write operations, enabling efficient updates, while the remaining structures are immutable and identical to columnar storage tables. Therefore, this study adopts a strategy of using a row-based storage structure in the in-memory table to support efficient updates and converting it to a columnar storage structure upon freezing to enable high-performance reads. Since the capacity of the in-memory table in an LSM-Tree is typically small, the size of the row-based storage structure can be effectively controlled. Figure~\ref{dasexstore:overview} illustrates the main components of \xstore, which include the following four key aspects:

\begin{enumerate}
    \item \textbf{Fine-Grained Row-to-Column Conversion:} The system consists of a top-level incremental row store and a column store. The incremental row store is used to support efficient data update operations. When the row store table reaches its capacity limit, it is frozen and awaits conversion to a column store table. Due to the small size of the row store table, fine-grained row-to-column conversion can be achieved.

    \item \textbf{Fine-Grained Compaction from Incremental Data to Baseline Data:} Considering the overhead of data compaction, \xstore introduces an intermediate layer between incremental data and baseline data. Incremental data is not directly compacted into the baseline data but is first merged into the intermediate layer, which consists of multiple column buckets. The system selectively compacts a portion of the data (one or more column buckets) into the baseline data at a time, enabling fine-grained compaction from incremental data to baseline data.

    \item \textbf{Fine-Grained Scheduling Strategy Based on Query Plans:} By implementing fine-grained row-to-column conversion and compaction operations, the scheduling strategy can select appropriate times to execute background tasks based on the current client query workload, thereby improving system stability. For columnar databases, query operations typically incur high overhead. This study finds that query plans can reflect the system's resource usage in the near future, allowing the estimation of idle periods for system resources.

    \item \textbf{Snapshot-Based Read Process:} Query operations in columnar databases often have long durations, and conflicts may arise between read operations and row-to-column conversion or compaction processes. To address this, this study implements a snapshot-based read process. At the start of a read operation, the system first acquires a data snapshot and then performs the read operation on this snapshot. Background update operations do not affect the acquired snapshot, thus avoiding conflicts between read operations and background tasks.
\end{enumerate}

\subsection{Overview of \xstore Design}

From a storage structure perspective, \xstore is divided into four layers. The first layer is the incremental row store, which stores newly inserted or updated data by users. When the row store table reaches its capacity limit, it is frozen and converted into a column store table, which is then placed in the incremental column store layer. Once the size of the incremental column store layer exceeds a predefined threshold, the data is compacted into the intermediate layer. When the capacity of a column bucket in the intermediate layer exceeds the threshold, the data is further compacted into the bottom-level baseline data layer. To effectively control the scheduling of row-to-column conversion and compaction tasks, \xstore's scheduler uses a cost model to estimate the overhead of user queries in the near future, thereby rationally scheduling background tasks. Below, we briefly describe the design of the column store table and the row store table.

\textbf{Columnar Storage Design:} To achieve fine-grained compaction and more flexible compaction strategies, the size of columnar tables in \xstore is limited to a predefined threshold (e.g., 4MB). Each columnar table internally adopts a columnar storage structure and stores data in primary key order. During query execution, the system only reads the required columns, thereby improving query efficiency. Notably, once a columnar table is constructed, it becomes immutable and does not support internal data modifications. To support delete operations, this study maintains a multi-version bitmap for each columnar table. This bitmap marks the validity of each row; when a bit in the bitmap is set to 0, it indicates that the corresponding row has been deleted.

\textbf{Row Storage Design:} To ensure that queries can quickly locate target rows and avoid the overhead of sorting primary keys during row-to-column conversion, this study uses an ordered structure to store incremental row data. Specifically, a skip list is chosen as the data structure for the incremental row store. Skip lists, as an efficient data structure, have been widely adopted in mature systems such as LevelDB~\cite{leveldb}, RocksDB~\cite{rocksdb}, and Redis~\cite{macedo2011redis}. Compared to common B+ tree in-memory structures, skip lists have a more compact memory layout and do not incur the overhead of node splitting, effectively reducing memory fragmentation. Thus, skip lists are particularly suitable for scenarios with frequent updates. However, before the row store table is frozen, since it needs to support insert and update operations, the position of each row is not fixed, making it impossible to use a bitmap to mark delete operations. Therefore, the row store table adopts an append-delete approach, where a delete marker is written into the row store table to mark the corresponding row as invalid when a deletion is required.

\section{Implementation Details of \xstore}

This section will elaborate on the implementation of basic operations in the \xstore system, the fine-grained row-to-column conversion and compaction mechanisms, and the specific implementation of the scheduling strategy.

\subsection{Implementation of Basic Operations in \xstore}

This section introduces the basic read, write, and update operations within the \xstore system. \xstore employs a multi-version concurrency control (MVCC) mechanism to handle concurrent operations. The implementation details of its read, write, update, and delete operations are described below.

\textbf{Insert Operation:} The insert operation first checks whether the new row already exists using the primary key clustered index. If it exists, the insertion fails; otherwise, the new row is inserted. Since columnar databases often perform bulk data imports, the system supports both batch and single-row insertions. For single-row or small-scale insertions, the new rows are written to the row store. For bulk insertions, the new rows are packed into a columnar table and placed in the incremental columnar storage section. The system maintains a globally auto-incremented version number, and each row is assigned a unique version number during insertion to support multi-version read operations.

\textbf{Multi-Version Read:} \xstore implements a snapshot-based read mechanism. During query execution, the read operation first acquires the latest system version number and the most recent system snapshot, and then performs the read operation on this snapshot. The read operation only reads data with a version number less than its own, ignoring other data. To efficiently acquire the system snapshot, \xstore maintains the latest system snapshot for quick access.

\textbf{Update Operation:} Update operations in columnar databases are typically implemented by marking old data for deletion and inserting new rows. When performing an update, \xstore locates the old row, marks it for deletion, and inserts the new row. Depending on the volume of data being updated, \xstore provides two storage methods for new rows: incremental columnar storage and incremental row storage. This study aims to ensure update efficiency while minimizing the impact on system performance. The implementation of these two update storage methods is described below.

Common update operations are executed via the \textbf{UPDATE} statement. The update implementation in \xstore includes the following steps:
(1) Locate the rows that meet the conditions using the \textbf{WHERE} clause;
(2) Insert new rows;
(3) Mark old rows for deletion.
If the volume of updated data is large (e.g., exceeding 4MB), the newly inserted rows are packed into a columnar table and added to the incremental columnar storage section to support efficient read operations. However, if the update involves only a few rows, packing them into a columnar table may cause memory fragmentation. To avoid generating a large number of small columnar tables, the system stores newly written rows in row storage when the update volume is small. Finally, \xstore marks the old rows for deletion. The following example illustrates this process. The update statement is:

\begin{lstlisting}[language=SQL]
UPDATE order 
SET c_comment = "Good" 
WHERE orderdate = '2024-09-20'
\end{lstlisting}

When executing the update operation, the \xstore system first traverses the orderdate column using a filter operator to locate all target rows. Then, the system calculates the size of the updated rows and decides the storage method based on whether the data volume exceeds a predefined threshold. If the update volume exceeds the threshold, the updated rows are packed into a columnar table and added to the incremental columnar storage section of \xstore. If the data volume is below the threshold, the updated rows are directly inserted into the row store. After completing the update, the system marks the old rows for deletion. Notably, the handling of delete marks differs between row and columnar tables. Since the filter operator has already determined the offset positions of the rows to be updated in the columnar table, the corresponding positions in the bitmap are simply set to 0. In contrast, rows in the row store lack fixed offset positions, so the row IDs of the invalid rows are recorded in the DList.

Considering the prevalence of single-row granularity updates in database operations, such as using Upsert for single-row updates, \xstore has been optimized accordingly. The system maintains a Bloom filter for each table to accelerate the search process. For example, in the following example, the system updates the status of the order with order\_id 123 to Shipped. The \xstore system searches top-down for the table (row or columnar) containing the row with order\_id 123. If a table does not contain the row, the system can skip searching that table using the Bloom filter, significantly improving search efficiency.

\begin{lstlisting}[language=SQL]
INSERT INTO orders (order_id, customer_id, order_status) 
VALUES (123, 1, 'Shipped') 
ON CONFLICT (order_id) DO UPDATE 
    SET order_status = EXCLUDED.order_status;
\end{lstlisting}

\textbf{Multi-Version Delete:} The \xstore system uses multi-version technology to read data and handles deletions in columnar or row tables by marking rows for deletion. However, this approach may lead to consistency issues in read operations. For example, when a query acquires a system snapshot to perform a read operation, if data that was initially readable is marked for deletion during the read process, and the newly inserted rows have a larger version number, the query may fail to read the data, resulting in inconsistency. Therefore, the system implements multi-version deletion marking, with different approaches for row and columnar tables. In row tables, deleted rows are appended to the row table, so only the version number of the update operation needs to be recorded in the delete mark. In columnar tables, a multi-version bitmap is designed to handle multi-version deletions. This bitmap consists of a version chain, where a new bitmap is appended to the chain for each delete operation. Each bitmap corresponds to a version number, allowing read operations to determine data visibility. To reduce the overhead of appending bitmaps for single-row deletions, the system records the offsets of deleted rows in a single delete mark version chain and applies them to the bitmap during reads. Additionally, as the version chain grows, the system releases old bitmaps when no active operations hold old version bitmaps to optimize resource utilization.

\subsection{Fine-Grained Row-to-Column Conversion and Compaction}

To minimize the potential impact of background operations on system performance, \xstore adopts fine-grained row-to-column conversion and compaction strategies, along with an efficient scheduling mechanism, significantly reducing the interference of individual background operations on system performance. The specific implementation of fine-grained row-to-column conversion and compaction is detailed below.

\textbf{Fine-Grained Row-to-Column Conversion:} This study proposes a fine-grained row-to-column conversion method to reduce the negative impact of row storage on read performance. Specifically, this method limits the size of row tables. When a row table reaches its capacity limit, it is frozen and added to a conversion queue. Frozen row tables no longer accept new data insertions, and the system creates a new row table to handle subsequent update and write requests. The scheduler converts these frozen row tables into columnar format at appropriate times to optimize storage efficiency and query performance.

\begin{figure}[t]
	\center		
	\subfigure[Traditional Compaction Process]{
		\label{dasexstore:compaction_conver}
		{\includegraphics[width=0.66\linewidth]{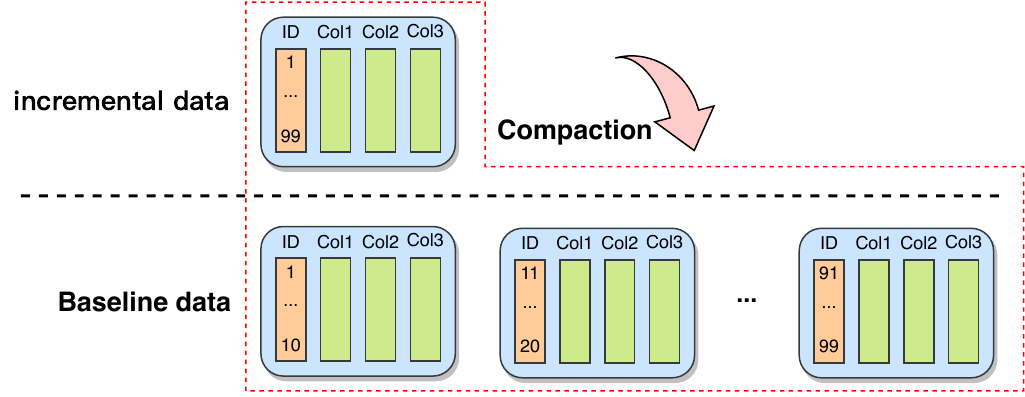}}
	}
	\subfigure[Compaction with Transition Layer]{
		\label{dasexstore:compaction_new}
		{\includegraphics[width=0.75\linewidth]{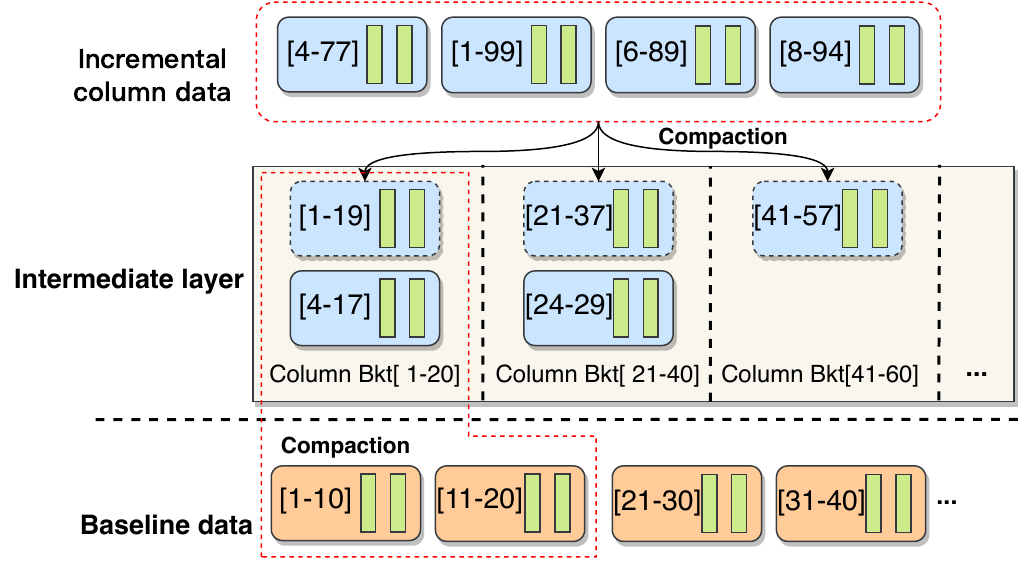}}
	}
	\caption{Design of the Transition Layer}
    \label{dasexstore:problem_update}
\end{figure} 

\textbf{Fine-Grained Compaction:} Due to the marking update feature of columnar databases, data is not actually released from storage when deleted. As update operations occur frequently, storage space usage continues to increase, necessitating periodic compaction of incremental data into baseline data. However, since client update operations are random, a single row or columnar table may cover a large amount of data. Therefore, when compacting incremental columnar data into baseline data, the overhead of a single compaction operation is significant, consuming substantial resources for foreground queries. The following example illustrates this. As shown in Figure~\ref{dasexstore:compaction_conver}, the range of an incremental columnar table covers the entire baseline data. The compaction task needs to read the entire baseline data and merge it with the incremental data, then pack the results into a columnar format for output. This process nearly rewrites the entire baseline data.

To address this issue, this study proposes a basic idea: horizontally partitioning the incremental columnar data into multiple intervals. This way, when compacting incremental data into baseline data, only one or a few intervals need to be merged, significantly reducing the overhead of a single compaction operation. As shown in Figure~\ref{dasexstore:overview}, to achieve partitioning, this study introduces a transition layer between the incremental columnar storage and the baseline data. This transition layer consists of multiple column buckets, where rows in different buckets do not overlap, and each bucket contains multiple columnar tables, allowing overlaps within the bucket. This design ensures that new data can be added to a column bucket without incurring compaction overhead. Since baseline data tables do not overlap, the range of column buckets is divided based on the range of baseline data, ensuring that each column bucket fully covers one or more baseline columnar tables. This design guarantees that the compaction of each column bucket with baseline data is conflict-free, facilitating concurrent compaction.

After introducing the transition layer, the compaction process in \xstore changes from directly compacting incremental data into baseline data to first compacting incremental data into the transition layer and then compacting the transition layer into baseline data. The following sections detail how these two compaction paths break down a single large-scale compaction into multiple small-scale compactions.

\textbf{Compaction from Incremental Data to the Transition Layer:} The traditional compaction process between incremental data and baseline data incurs significant overhead because it requires handling both incremental and baseline data, with baseline data typically being large in scale. To address this issue, directly compacting incremental columnar data into the transition layer is an effective strategy. However, as the data volume in the transition layer grows, the compaction overhead also increases, reintroducing the original problem. Therefore, when compacting incremental data into the transition layer, it is not merged with existing data in the transition layer but is directly stored in the transition layer, as shown in Figure~\ref{dasexstore:compaction_new}. The specific compaction process is as follows: (1) The selected incremental columnar data is merge-sorted by the primary key, and invalid data is removed; (2) The merged data is packed into a new columnar table and stored in the column bucket of the transition layer. Since the size of columnar tables is limited by a threshold and the data range of columnar tables within a bucket must strictly adhere to the bucket's range, the creation of a new columnar table stops when its size exceeds the threshold or reaches the bucket boundary, and a new columnar table is created to continue writing; (3) The newly generated columnar table is placed in the corresponding column bucket. Thus, the size of the compaction from incremental data to the column bucket depends only on the size of the input incremental data, as calculated in Formula~\ref{equation:ct}.

\begin{equation}\label{equation:ct}
C_{t} = \sum_{i \in \Omega} s_i
\end{equation}

Here, $C_t$ represents the size of a single compaction, $\Omega$ represents the set of input columnar tables for the compaction, and $S_i$ represents the size of the $i$-th columnar table in the set. The scheduler controls the compaction granularity by managing $\Omega$, where the sum of the sizes of the columnar tables in the set should be less than or equal to $G$, representing the granularity of a single compaction.

\textbf{Compaction from the Transition Layer to Baseline Data:} When the data volume in a column bucket exceeds a certain threshold, it needs to be compacted into the baseline data. The compaction process includes reading the column bucket data and baseline data, merge-sorting them, removing expired data, and packing the results into a columnar table to be written into the baseline data layer. \xstore sets a threshold $T$ for column buckets. When the data volume in a column bucket exceeds this threshold, the compaction operation is triggered. Since the range of a column bucket fully covers one or more baseline data files, compaction operations between different column buckets do not cause data conflicts. This allows the scheduling strategy to flexibly select one or more column buckets for compaction when needed, achieving free control over compaction granularity. The size of the compaction from the $i$-th column bucket to the baseline data, $C_{i}$, is calculated as in Formula~\ref{equation:xstore:ci}.

\begin{equation}\label{equation:xstore:ci}
C_{i} = \sum_{j \in \Gamma_i}s_j + \sum_{k \in \beta_i}s_k
\end{equation}

In this section, $\Gamma_i$ represents the set of columnar tables in the $i$-th column bucket, where $S_j$ represents the size of the $j$-th columnar table in the set. $\beta_i$ represents the set of baseline columnar tables corresponding to the $i$-th column bucket, where $S_k$ represents the size of the $k$-th columnar table in the set. The range of $\beta_i$ is determined by the range of the corresponding column bucket, and the size of the data in $\Gamma_i$ is determined by the threshold $T$. In short, the size of the compaction from a column bucket to baseline data depends on the size of the data in the column bucket and the size of the corresponding baseline data.

Next, we calculate the size of a single compaction $C$ without using column buckets. Assuming the range of incremental data is random and covers the entire baseline data, as shown in Formula~\ref{equation:xstore:cc}.

\begin{equation}\label{equation:xstore:cc}
C = \sum_{i \in \Omega} s_i + \sum_{i=1}^n\sum_{j \in \Gamma_i}s_j + \sum_{i=1}^n\sum_{k \in \beta_i}s_k = C_t + \sum_{i=1}^nC_i
\end{equation}

Here, $n$ is the number of column buckets. The data volume of the traditional compaction process is the sum of all incremental data and baseline data, with a compaction granularity much larger than that of compaction from incremental data to the transition layer or from the transition layer to baseline data. As shown in the above formula, the introduction of the transition layer significantly reduces compaction granularity.

\textbf{Management of Column Buckets:} The transition layer consists of multiple column buckets. Once the data in a column bucket is compacted into the baseline data, the volume of baseline data corresponding to that column bucket increases. As shown in the formula for $C_{i}$, the compaction from a column bucket to baseline data depends on the size of the column bucket and the size of the baseline data. As the baseline data increases, the compaction granularity from the column bucket to baseline data gradually becomes coarser. To control this issue, the system decides whether to split a column bucket based on the size of the baseline data corresponding to the column bucket, as determined by Formula~\ref{equation:xstore:Split}.

\begin{equation}\label{equation:xstore:Split}
Split(i) = G - T - \sum_{k \in \beta_i}s_k
\end{equation}

When $Split(i) < 0$, it indicates that the baseline data corresponding to the current column bucket is excessive, and a split operation is required. At this point, the column bucket is split into two independent column buckets, each covering half of the baseline data corresponding to the original column bucket and covering complete files. This way, the compaction overhead for each column bucket is effectively controlled.

\textbf{Concurrency Control:} Background compaction tasks face two main challenges when executed concurrently: First, multiple compaction tasks may attempt to compact the same set of tables simultaneously, causing contention; second, changes to tables during compaction may lead to concurrency conflicts, such as a columnar table being read by a user query being deleted during compaction. To address these issues, this study proposes corresponding solutions.

The contention among multiple compaction tasks mainly arises from two aspects: First, incremental columnar data is compacted into the transition layer; second, data in the transition layer is compacted into baseline data. To address these issues, this study designs the following two mechanisms:
(1) Before executing a compaction task, the system sets a compaction mark on the target table. Once a table is marked as being compacted, subsequent compaction tasks cannot operate on it, effectively avoiding contention for incremental columnar data compaction.
(2) By ensuring that the ranges of column buckets do not overlap and that the baseline data covered by each column bucket does not overlap, conflict-free compaction between column buckets is achieved. Additionally, the compaction from baseline data to column buckets adopts an append-write approach for new columnar tables and uses a simple producer-consumer model to synchronize the two compaction tasks.

\begin{figure}
    \centering
    \includegraphics[width=0.75\linewidth]{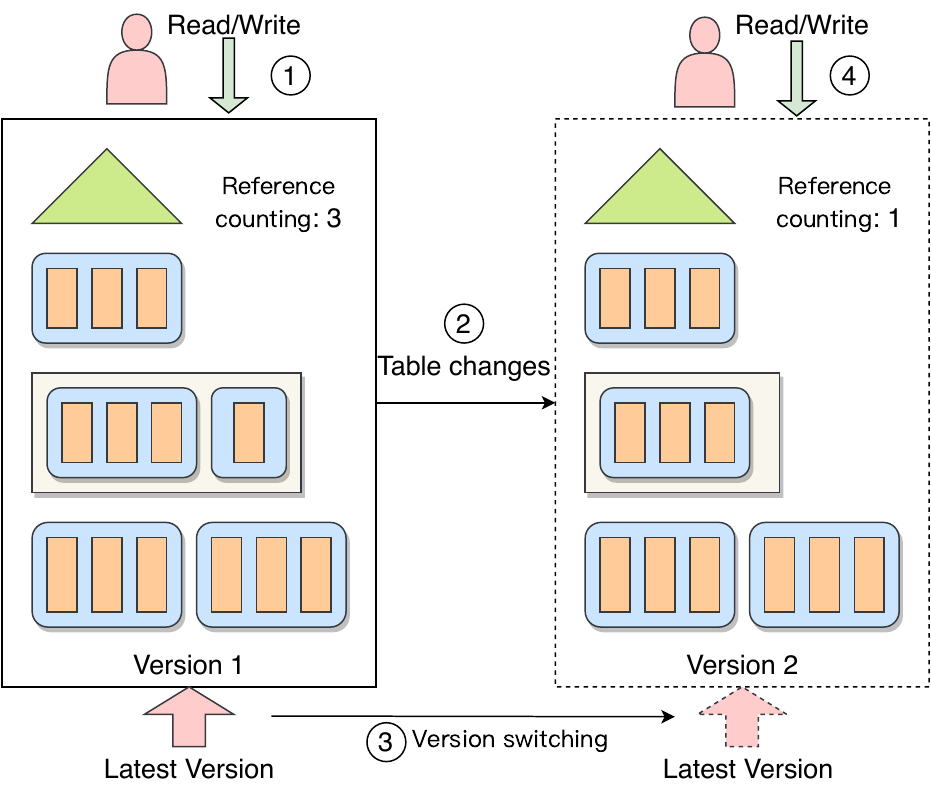}
    \caption{Multi-Version Concurrency Control}
    \label{dasexstore:snapshot}
\end{figure}

To avoid conflicts caused by table modifications during compaction with other operations, this study employs a multi-version management mechanism, as shown in Figure~\ref{dasexstore:snapshot}. The \xstore system maintains the latest version of the system, where each version is equivalent to a snapshot of all tables in \xstore, storing all tables visible to users at that time. When users access \xstore, their operations are based on the current latest version (\ding{172}). When row-to-column conversion or background compaction causes table changes, \xstore generates a new version in the background (\ding{173}). After the new version is generated, the system updates the pointer to the latest version to the newly generated version (\ding{174}). Thereafter, all user accesses are based on the latest version (\ding{175}). Considering that there may be unfinished user operations during version switching, each version maintains a reference count, and the version is only released when the reference count is 0, ensuring that unfinished read operations can continue to execute.

\subsection{Scheduling Strategy}

Compared to traditional row-store databases, the queries processed by columnar databases are typically more complex and incur higher overhead. However, in actual system operation, resource utilization is not always at 100\%, and there are still relatively idle periods. \xstore implements fine-grained row-to-column conversion and compaction operations, and its background task scheduling mechanism is more flexible. The scheduling goal of \xstore is to fully utilize background resources while minimizing the impact of background tasks on foreground operations. The scheduling strategy needs to address two key issues: (1) When to schedule background tasks; (2) Which background tasks to schedule.

\textbf{Choosing the Timing for Scheduling Background Tasks:} In computer systems, the number of CPU cores ($N$) is typically limited, which restricts the number of tasks the system can handle in parallel. Therefore, for the \xstore scheduler, the number of tasks occupying the CPU at the same time ($t$) should not exceed $N$. These tasks include user-initiated query operations ($q$) and background tasks ($g$). In database systems, user query operations usually have higher priority than background tasks, so the scheduler should prioritize meeting the resource demands of foreground query tasks and schedule waiting background tasks during idle periods. However, due to the randomness of user requests, if background tasks occupy all cores, subsequent user requests may experience blocking. Previous research often reserves idle resources for foreground tasks or suspends background tasks to avoid this issue, but these methods introduce new challenges. Reserving resources may lead to resource waste, while suspending background tasks incurs significant scheduling overhead.

\begin{figure}
    \centering
    \includegraphics[width=0.5\linewidth]{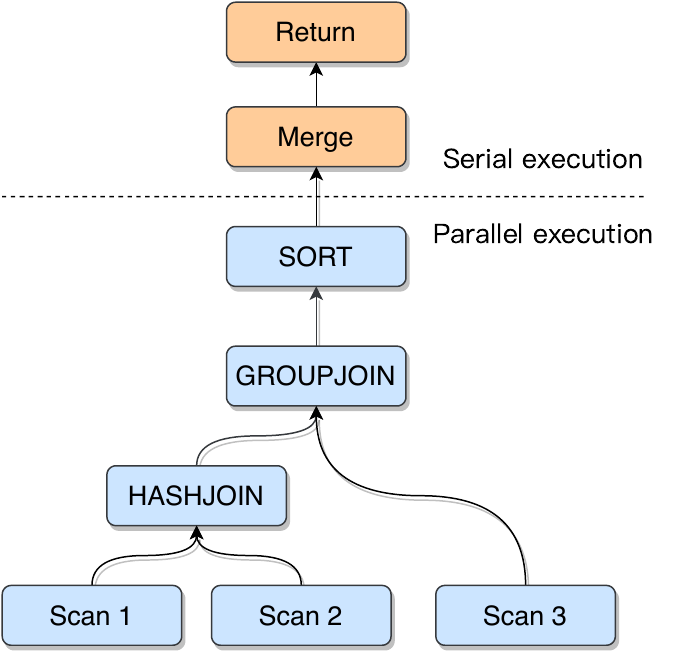}
    \caption{Operator Execution Process}
    \label{dasexstore:parallel}
\end{figure}

This study proposes a new solution: estimating the resource usage of query plans for a future period based on the user's query plan and scheduling background tasks accordingly. In columnar databases, the construction of query plans typically relies on a cost model, which estimates the execution cost of different query plans and selects the optimal one for execution. Using the cost model and execution plan, the resource usage of queries for a future period can be estimated. As shown in Figure~\ref{dasexstore:parallel}, the query plan consists of multiple operators, some of which are executed serially while others are executed in parallel, with the degree of parallelism varying over time. During periods of low parallelism (i.e., low CPU resource usage), the system has idle core resources, allowing background tasks to be scheduled.

However, this approach also faces challenges. The cost model's estimates are often inaccurate. For example, the cost model may estimate the execution time of a Scan operator as 0.5 seconds, while the actual execution time may be 1 second. This inaccuracy can lead to incorrect scheduling, potentially causing excessive scheduling of background tasks and crowding out resources for foreground tasks. To reduce estimation errors, this study introduces a constant $\phi_i$ for each operator $i$'s cost model result $Cost_i$ to correct the error between the cost model and the actual execution time. Therefore, the scheduler calculates the execution time $Duration_i$ of each operator using Formula~\ref{equation:xstore:Duration_i}.

\begin{equation}\label{equation:xstore:Duration_i}
Duration_i = Cost_i * \phi_i
\end{equation}

In this study, $Cost_i$ represents the execution time estimated by the cost model. Next, we will detail how to calculate the constant $\phi_i$. The constant $\phi_i$ is used to correct the error between the cost model and the actual execution time, so $\phi_i$ is calculated based on the actual execution time and the cost model's estimated overhead for each operator. However, considering performance fluctuations, even the same operator executing the same data may have varying execution times. Therefore, $\phi_i$ is continuously updated as queries are executed. However, simply assigning a new value to $\phi_i$ each time is unreasonable, so $\phi_i$ is calculated as the average of the past actual execution times and cost model estimates. To reduce the overhead of calculating the average of $\phi_i$, this study adopts the \textit{Welford algorithm} to compute the average of $\phi_i$, using Formula~\ref{equation:xstore:phi_new} and Formula~\ref{equation:xstore:phi_i}.

\begin{equation}\label{equation:xstore:phi_new}
\phi_i^{new} = \phi_i^{'} + \frac{\phi_i^{'} - \phi_i^{old}}{n} 
\end{equation}

\begin{equation}\label{equation:xstore:phi_i}
\phi_i^{'} = \frac{Duration_i^{'}}{Cost_i}
\end{equation}

In the formulas, $\phi_i^{new}$ represents the updated value of $\phi_i$, $\phi_i^{old}$ represents the previous value of $\phi_i$, $n$ represents the number of updates, and $\phi_i^{'}$ represents the ratio of the actual execution time $Duration_i^{'}$ to the cost model's estimated overhead $Cost_i$ for the current query. After each query execution, the corresponding operator's $\phi_i$ is updated. As queries are executed, the system's estimated execution time becomes increasingly accurate.

Although this study can more accurately estimate the execution time of each operator during query execution, concurrently executed operators may deviate from the expected execution process due to various reasons. For example, if new query requests arise during execution, they may occupy CPU cores that were expected to be idle in the near future, causing background tasks scheduled to run during that period to be delayed. To address this, this study designs a monitoring thread for \xstore to track the execution process of currently running queries in real time and adjust the scheduling of background tasks accordingly. To minimize the resource consumption of the monitoring thread, it is awakened periodically (e.g., every 100 milliseconds).

\textbf{Selecting Background Tasks to Schedule:} The previous section discussed when resources become idle for scheduling background tasks. Next, the scheduler must determine which background tasks to schedule. Intuitively, row-to-column conversion tasks have higher priority than background compaction tasks because, as shown in the experimental results in Figure~\ref{dasexstore:impact_update}, the impact of incremental row storage on performance is greater than that of incremental columnar storage. Therefore, converting row storage to columnar format in a timely manner is more urgent. Thus, when the system has idle resources in the near future, row-to-column conversion tasks are prioritized for scheduling.

\section{Experiments}

This section aims to evaluate the performance of the \xstore system, focusing on the performance improvements brought by fine-grained row-to-column conversion, the impact of the cost-based fine-grained compaction strategy on system stability, and the performance comparison of \xstore with other systems in mixed workload scenarios.

\subsection{Experimental Setup}

\textbf{Comparison Systems:}
(1) **DuckDB**: DuckDB~\cite{raasveldt2019duckdb} is a lightweight embedded analytical database management system (DBMS) developed by the Centrum Wiskunde Informatica (CWI) in the Netherlands. It uses a columnar storage structure and has gained widespread support in the open-source community. Its design goal is to provide efficient SQL query execution while maintaining ease of use and low resource consumption.
(2) **TiDB**: TiDB~\cite{tidb} is a relational database that supports HTAP workloads. Its storage system is based on the Log-Structured Merge-Tree (LSM-Tree). It provides online transaction processing (OLTP) capabilities through row storage and analytical processing (OLAP) capabilities through columnar storage, achieving efficient integration of transactional and analytical workloads.
(3) **\xstore**: \xstore is the columnar storage-based LSM-Tree storage system developed in this study, with a codebase exceeding 20,000 lines. It integrates the design philosophy of LSM-Tree and incorporates the optimization techniques proposed in this study. By combining the advantages of incremental row storage and columnar storage, \xstore achieves efficient update operations and optimized query performance. Specifically, the system uses incremental row storage to support real-time updates and converts row storage tables to columnar storage upon freezing to improve query efficiency. Additionally, \xstore balances update and query performance through fine-grained row-to-column conversion and compaction strategies, as well as a query plan-based scheduling mechanism, minimizing the interference of background operations on foreground queries. The system also integrates multi-version read and concurrency control mechanisms to ensure data consistency and real-time performance.

\textbf{Testing Tools:}
(1) **XBench**: This study developed the XBench benchmark tool, which supports user-defined test workloads and allows flexible configuration of the proposed optimization methods (e.g., fine-grained row-to-column conversion, fine-grained compaction, and cost-based compaction models) during evaluation, enabling accurate assessment of the performance improvements of different optimization strategies.
(2) **Mixed Workload**: The mixed workload (transactional/analytical processing) is a composite workload that supports both transactional operations (e.g., insert, update, delete) and analytical queries. This study implemented the test environment for this workload based on previous work~\cite{LASER}.

\textbf{Experimental Environment:}
All experiments were conducted on a test machine equipped with two Intel(R) Xeon(R) Gold 6240R 2.40GHz processors and 128 GB of memory. A 1TB INTEL SSDPE2KE032T8 NVMe SSD was used as the storage device. The operating system was CentOS 7.8.

\subsection{Performance Optimization of Fine-Grained Row-to-Column Conversion}

This subsection focuses on evaluating the performance improvements of the fine-grained row-to-column conversion strategy on system update and query performance.

\begin{figure}
    \centering
    \includegraphics[width=0.6\linewidth]{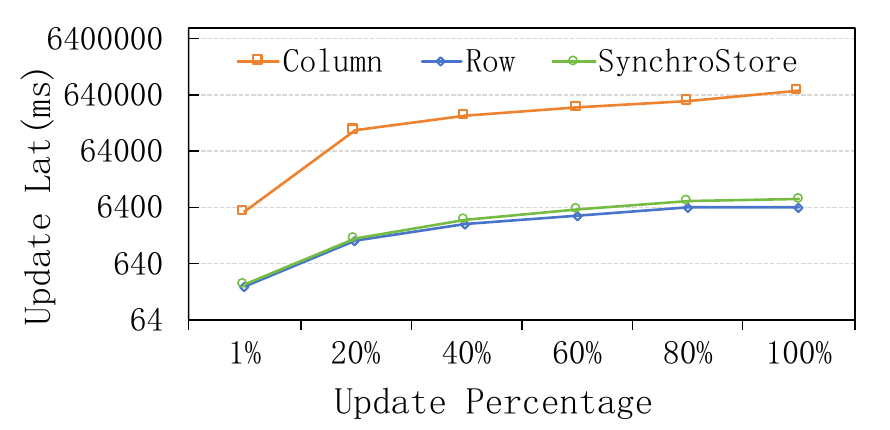}
    \caption{Performance Improvement of Fine-Grained Row-to-Column Conversion on Update Performance}
    \label{figs:DaseXStore:exp1}
\end{figure}

\textbf{Improvement in Update Performance:}
The experiment compares the impact of fine-grained updates on update performance under different update ratios. The experiment first inserts 10GB of data and then performs update operations at different ratios. Updates are performed using random single-row updates (i.e., Upsert), updating only one column at a time, with each row containing 30 columns. The time taken to complete these update operations is compared. Three configurations were used:
(1) **Incremental Columnar**: Uses columnar format as the incremental storage format, friendly to query performance, based on \xstore.
(2) **Incremental Row**: Uses row storage format as the incremental storage format, theoretically optimal for update performance but unfriendly to query performance, based on \xstore.
(3) **\xstore**: Uses row storage for incremental updates and implements fine-grained row-to-column conversion.
Figure~\ref{figs:DaseXStore:exp1} shows the performance of the three configurations under different update ratios. The results show that when the update ratio is 1\%, the update overhead of \xstore is only 4.8\% of that of incremental columnar updates and only about 5\% higher than that of incremental row updates. As the update ratio increases, the time required to complete updates also increases. When the update ratio reaches 100\%, the update overhead of \xstore is only 1.2\% of that of incremental columnar updates. This is attributed to the efficient handling of update requests by the row storage structure of \xstore, without introducing excessive additional overhead.

\begin{figure}[h]
	\centering
    \subfigure[Query Performance Under Different Update Ratios]{
		\label{figs:DaseXStore:exp1.2} 
		\includegraphics[width=0.45\linewidth]{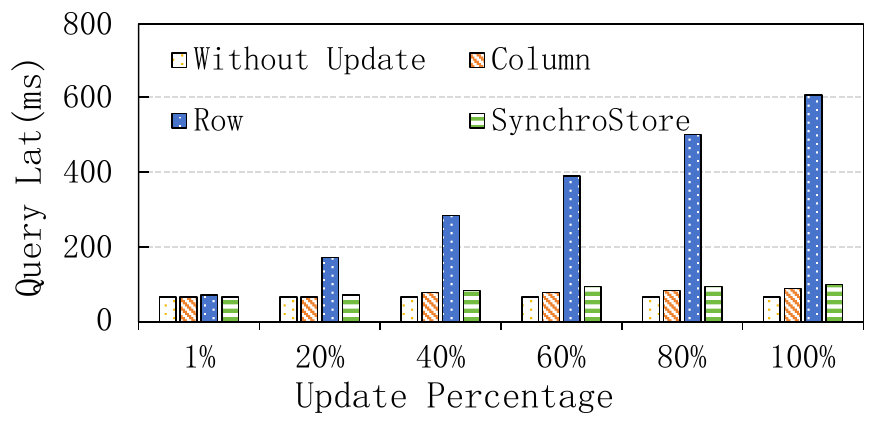}
	}
	\subfigure[Query Performance Under Different Projection Sizes]{
		\label{figs:DaseXStore:exp1.3}
		\includegraphics[width=0.45\linewidth]{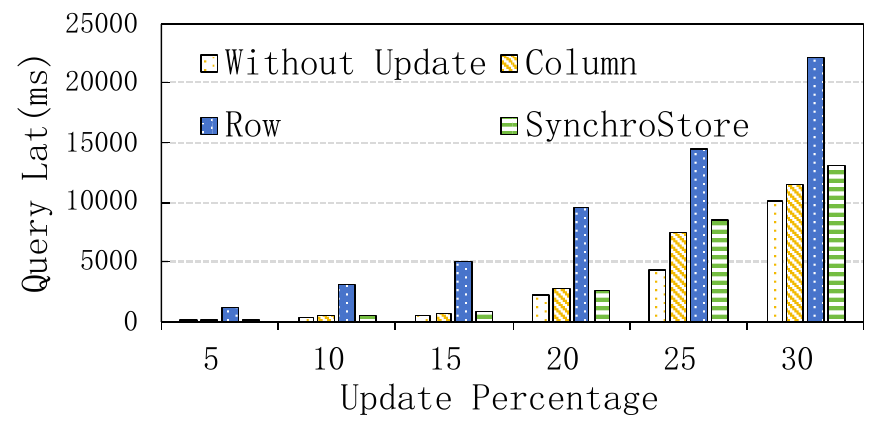}
	}
	\caption{Performance Improvement of Fine-Grained Row-to-Column Conversion on Query Performance}
	\label{figs:DaseXStore:exp1.3} 
\end{figure} 

\textbf{Impact on Query Performance:}
This experiment evaluates the impact of fine-grained row-to-column conversion on query performance under different update ratios. The experiment first imports 10GB of data and then performs update operations at different ratios. Updates are performed using random single-row updates (i.e., Upsert), with each query involving only one column. Each row contains 30 columns, consisting of string and integer types. The query performance of different configurations after updates is compared. Four configurations were used:
(1) **No Updates**: \xstore directly executes queries after importing data, reflecting the system's optimal query performance.
(2) **Incremental Columnar**: Uses columnar format as the incremental storage format, friendly to query performance, based on \xstore.
(3) **Incremental Row**: Uses row storage format as the incremental storage format, theoretically optimal for update performance but unfriendly to query performance, based on \xstore.
(4) **\xstore**: Uses row storage for incremental updates and implements fine-grained row-to-column conversion. Only the fine-grained row-to-column conversion optimization technique is used in this experiment.
Figure~\ref{figs:DaseXStore:exp1.2} shows the impact of the four configurations on query performance under different update ratios. The results show that when the update ratio is 20\%, the query latency of \xstore is 2\% higher than that of incremental columnar storage but 57\% lower than that of incremental row storage. As the update ratio increases, the query performance of all configurations is affected, but incremental columnar storage and \xstore are less affected. When the update ratio reaches 100\%, the query latency of \xstore is only 15\% of that of incremental row storage, because fine-grained row-to-column conversion efficiently converts incremental row storage to columnar storage, avoiding the negative impact of incremental row storage on performance.

Considering the significant impact of query projection size on query performance, the experiment further tests the query performance of the configurations under different projection sizes. The experiment first inserts 10GB of data and then performs update operations on all inserted data using random single-row updates (i.e., Upsert), with each row size approximately 4KB. The query performance of the four configurations after updates is tested. Figure~\ref{figs:DaseXStore:exp1.3} shows the read performance of the four configurations under different projection sizes. When the projection size is 5, the query latency of \xstore is only 16\% of that of incremental row storage and about 3\% higher than that of incremental columnar storage. As the projection size increases, the performance gap between the four configurations gradually narrows, because larger projections require reading more data.

\subsection{Overhead of Fine-Grained Compaction}

Fine-grained compaction (including fine-grained row-to-column conversion and fine-grained compaction) significantly reduces the overhead of a single compaction operation and provides more optimization space for scheduling strategies. This subsection tests the overhead of fine-grained compaction.

\begin{figure}
    \centering
    \includegraphics[width=0.85\linewidth]{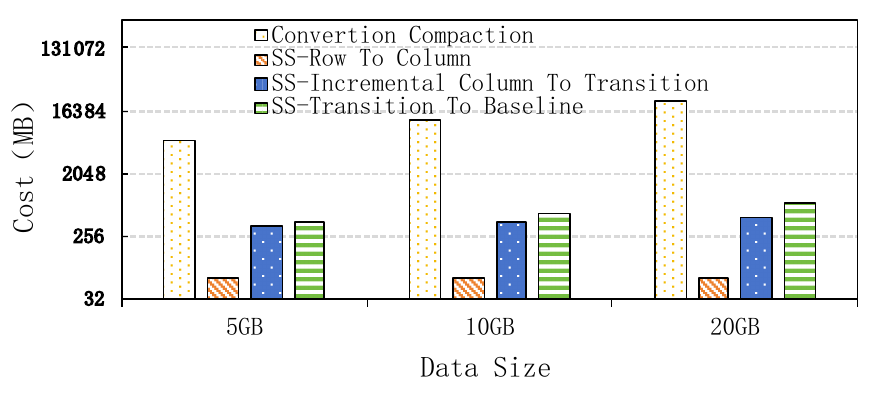}
    \caption{Compaction Overhead}
    \label{figs:DaseXStore:exp2.1}
\end{figure}

This experiment evaluates the overhead comparison between fine-grained compaction mechanisms and traditional compaction mechanisms under different data volumes. The experiment first imports datasets of different sizes and then performs update operations on all data to measure the average overhead of various compaction operations. Four compaction operations are tested, including traditional compaction, row-to-column conversion, and compaction operations. The specific compaction processes are as follows:
(1) **Traditional Compaction**: Based on a simple threshold mechanism, when incremental storage reaches a preset threshold, the system directly compacts incremental data into baseline data. Due to the randomness of incremental data, this process often incurs significant overhead.
(2) **SS-Row-to-Column Conversion**: The fine-grained row-to-column conversion operation of \xstore. When the incremental table in row storage format reaches the threshold, it is converted to columnar format and stored as incremental columnar data.
(3) **SS-Incremental Columnar to Transition Layer**: The compaction operation from incremental columnar storage to the transition layer in \xstore. This operation selects several incremental columnar tables to merge and place into the transition layer, avoiding direct compaction with baseline data.
(4) **SS-Transition Layer to Baseline Data**: The compaction operation from the transition layer to baseline data in \xstore. This operation selects a column bucket to compact into baseline data, releasing data marked for deletion.
Figure~\ref{figs:DaseXStore:exp2.1} shows the average overhead of each compaction operation under different data volumes (unit: MB). The results show that as the data volume increases, the overhead of traditional compaction grows linearly. When the system data volume exceeds 20GB, the overhead of a single compaction operation also exceeds 20GB. In contrast, the compaction overhead of \xstore is effectively controlled. The overhead of SS-Row-to-Column Conversion depends only on the size of the incremental row storage table, which is set to 64MB in this experiment, so its overhead is constant at 64MB. As the data volume increases, the overhead of SS-Incremental Columnar to Transition Layer and SS-Transition Layer to Baseline Data increases, but the growth rate is much lower than the increase in data volume. The maximum overhead of a single compaction is 795MB, significantly lower than that of traditional compaction. This optimization is mainly due to the introduction of the transition layer, which avoids direct compaction of incremental data into baseline data, thereby greatly reducing the overhead of a single compaction operation.

\subsection{Performance Testing Under Mixed Workloads}

This subsection evaluates the performance of the proposed optimization scheme in mixed workload scenarios, focusing on the performance improvement of the cost-based scheduling mechanism on query performance and comparing it with existing systems.

\textbf{Mixed Workload:}
The mixed workload design in this experiment refers to the benchmark workload proposed in previous research~\cite{LASER}. The benchmark includes common transactional and analytical queries, consisting of 5 SQL statements designed to simulate typical operations in real database environments. These operations cover basic operations such as data insertion, query, update, and aggregation, comprehensively evaluating the performance of database systems in handling mixed workloads. Specifically, the SQL statements are designed as follows:

\begin{enumerate}[label=\arabic*]
    \item (SQL1) Insert a new row, simulating data write operations;
    \item (SQL2) Update specific columns of a row, simulating data modification in transactional operations;
    \item (SQL3) Perform a sum operation on selected columns of a row, simulating aggregation operations in analytical queries;
    \item (SQL4) Calculate the maximum value of selected columns on a selected row, testing the database's performance in complex queries;
    \item (SQL5) Perform a join query on two tables, combined with conditional filtering, group aggregation, and sorting operations, simulating the database's performance in complex analytical queries.
\end{enumerate}

These queries are written in SQL as follows:

\begin{lstlisting}[language=SQL]
(SQL1) INSERT INTO table_1 
     VALUES (col_0, col_1, ..., col_{c}

(SQL2) UPDATE table_1 
     SET col_1 = v_1,...,col_k = v_k
     WHERE col_0 = v

(SQL3) SELECT col_1 +col_2 +...+col_{n} 
     FROM table_1 
     WHERE col_0 in [v_start,v_end)

(SQL4) SELECT MAX(col_1), ..., MAX(col_k) 
     FROM table_1 
     WHERE col_0 in [v_start,v_end)

(SQL5) SELECT t1.col_0, t1.col_1
     FROM table_1 t1
     JOIN table_2 t2 ON t1.col_1 = t2.col_0
     WHERE t1.col_2 in [v_start, v_end)
\end{lstlisting}

In the experimental design, parameters $k$, $v_k$, $v_start$, and $v_end$ are used to control the projection size and query range of the query statements. The table structure used in the experiment contains 31 columns, where col\_0 is the integer primary key, and the remaining 30 columns (col\_1, col\_2, ..., col\_30) consist of integer and string types.
The experimental process mainly includes the following steps: First, data is loaded into the database, and then the workload is executed on this dataset. During the execution, the system records key performance indicators, including response time, throughput, and resource usage. By comparing and analyzing these performance data, this study can evaluate the performance differences of different database storage engines under mixed workloads.
It is important to note that this experiment does not use complex OLAP queries from benchmarks such as TPC-H and TPC-DS, as the research focus is on evaluating the performance of storage engines under mixed workloads rather than query optimization strategies. To ensure the fairness and repeatability of the experiment, all queries are executed under the same dataset and configuration conditions, and the performance differences under different workloads are systematically analyzed.

\textbf{Performance Improvement of Cost-Based Scheduling:}
Although fine-grained compaction techniques can alleviate resource contention caused by background compaction operations to some extent, their effectiveness is still limited. To further optimize system performance, this study proposes and tests a cost-based scheduling strategy. The experiment uses a mixed workload testing method, first importing 20GB of data and then executing the mixed workload test. Two system configurations were used:
(1) **\xstore**: Deploys all optimization techniques proposed in this study.
(2) **\xstore-NoScheduler**: \xstore without the cost-based scheduling mechanism.
Table~\ref{tab:dasxstore:taillatency} shows the tail latency of Q1 queries under the mixed workload for the two configurations. It can be seen that at the 75th percentile, the cost-based scheduling model reduces tail latency by about 20\%. As the percentile increases, the performance improvement of the cost-based scheduling strategy becomes more significant, reducing tail latency by up to 34\%. This is because the cost-based scheduling model alleviates resource contention caused by background tasks.

\begin{table}[t]
\centering
\begin{tabular}{llllll}
\hline
\multicolumn{1}{c}{}   & P50   & P75   & P99    & P99.9  & P99.99  \\ \hline
\xstore              & 35.9  & 59.44 & 261.33 & 364.43 & 475.24 \\
\xstore -NoScheduler & 39.58 & 73.32 & 356.69 & 553.3  & 731.22 \\ \hline
\end{tabular}
\caption{Tail Latency Under Mixed Workload}
\label{tab:dasxstore:taillatency}
\end{table}

\begin{figure}[t]
	\centering
    \subfigure[Performance of Insert and Update Operations Under Mixed Workload]{
		\label{figs:DaseXStore:exp3.1} 
		\includegraphics[width=0.42\linewidth]{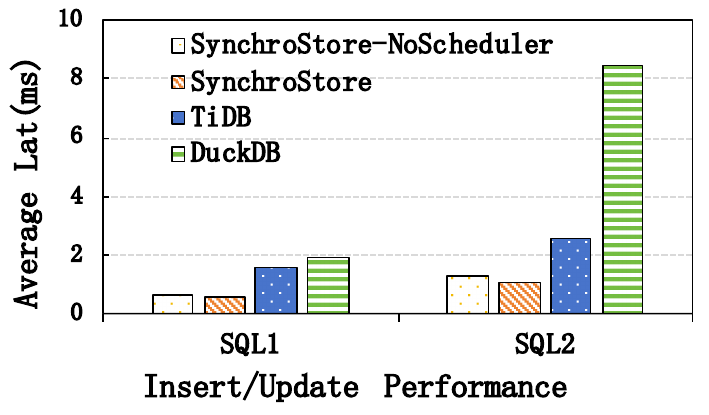}
	}
	\subfigure[Performance of Query Operations Under Mixed Workload]{
		\label{figs:DaseXStore:exp3.2}
		\includegraphics[width=0.52\linewidth]{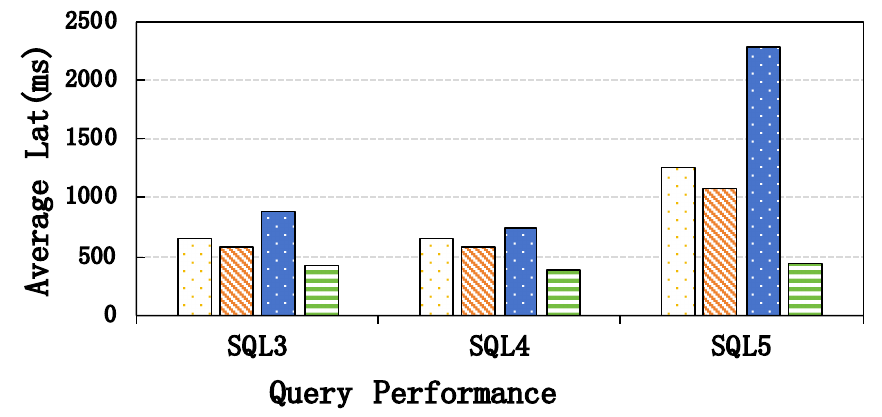}
	}
	\caption{Insert, Update, and Query Latency of Different Systems Under Mixed Workload}
	\label{figs:DaseXStore:exp3} 
\end{figure} 

\textbf{Query Latency:}
This experiment compares the performance of different systems under mixed workloads using the mixed workload test. The experiment first imports 20GB of data and then runs the mixed workload. Three configurations were used:
(1) DuckDB: DuckDB with default configuration.
(2) \xstore: Deploys all optimization techniques proposed in this study.
(3) \xstore -NoScheduler: \xstore without the cost-based scheduling mechanism.
Figure~\ref{figs:DaseXStore:exp3} shows the query latency of the three configurations under the mixed workload. Figure~\ref{figs:DaseXStore:exp3.1} shows the insert and query latency of the three systems. It can be seen that \xstore has significant advantages over TiDB and DuckDB in both insert and update operations. The insert latency of \xstore is 32\% and 27\% of that of TiDB and DuckDB, respectively, and the update latency is 41\% and 12\% of that of TiDB and DuckDB, respectively. Meanwhile, \xstore with the cost-based fine-grained scheduling mechanism has 17\% and 19\% lower insert and update latency than \xstore -NoScheduler. This is because \xstore uses incremental row storage to handle updates, resulting in better insert and update performance. Additionally, fine-grained row-to-column conversion and the cost-based scheduling strategy reduce the impact of incremental row storage on read performance and the impact of the compaction process on performance.

\section{Related Work}

In recent years, with the increasing demand for data analysis, more and more attention has been paid to the query performance of databases. As a result, the columnar storage format has gained significant attention. However, in mixed workloads, databases not only need to perform efficient query operations but also support efficient updates and inserts. Considering that columnar formats often use specialized encoding schemes, making it difficult to update once constructed, most columnar storage databases adopt incremental updates. Updated data is written to the incremental data area and later synchronized with the original baseline data. Below, we introduce the incremental updates in columnar storage systems and several common synchronization methods.

\section{Incremental Updates in Columnar Storage:}

One challenge of columnar updates is the inability to perform in-place updates easily. To ensure high space utilization efficiency and improve query performance, data in columnar engines is encoded using various methods such as run-length encoding, dictionary encoding, bit-vector encoding, prefix encoding, and delta encoding. To modify this encoded data, it must first be decoded, updated, and then re-encoded and written back. This not only results in significant update overhead but also causes write amplification, increasing disk wear. Therefore, modern columnar storage systems~\cite{schulze2024clickhouse, raasveldt2019duckdb, lipcon2015kudu, Doris, adb, tidb} almost universally adopt incremental updates. Below, we introduce several typical columnar storage systems.

\textbf{ClickHouse}~\cite{schulze2024clickhouse} is currently a popular columnar storage system. Its storage structure, called MergeTree, resembles a Log-Structured Merge-Tree (LSM-Tree) without a Memtable. Updates are implemented through append writes and compaction. ClickHouse maintains a flag for each row: -1 indicates deletion, and 1 indicates insertion. During an update, the old row is read, marked as -1, and appended to the database, followed by a new row marked as 1. During queries, invalid records are discarded, and only the final valid rows are retained. This approach converts updates into append writes, avoiding the overhead of directly modifying data blocks, but requires the application layer to handle state filtering logic during queries. However, ClickHouse packs even a single updated row into columnar format, which can easily lead to significant fragmentation.

\textbf{Apache Kudu}~\cite{lipcon2015kudu}, a distributed columnar storage engine for real-time analytics, is particularly friendly to updates, especially single-row updates. Its storage structure resembles an LSM-Tree with only the base level. Kudu supports column-level updates. If only a single column of a row is updated, Kudu first queries the row number based on the primary key, records the row number and the updated column value in the in-memory DeltaMemStore, and flushes it to disk when a threshold is reached, later merging it with the original data. Since Kudu does not insert the entire row during updates, old data is invalidated at the column level, reducing space waste. However, during reads, Kudu must read both the original data and the column updates to reconstruct the complete row, which affects read performance.

Additionally, many columnar storage systems~\cite{Doris, adb, tidb} adopt LSM-Tree-like storage structures. The basic principle of their updates is to mark old rows for deletion, insert new rows into the incremental data section, and later synchronize the incremental data with the baseline data. Although this design is more update-friendly, it introduces synchronization overhead. Thus, how to synchronize incremental data with baseline data becomes a critical issue.

\section{Synchronization in Columnar Engines:}

Current synchronization techniques can be summarized into three categories: two-phase migration, dictionary-sorted merge-based synchronization, and threshold-driven synchronization. Below, we introduce each.

\textbf{Two-Phase Migration in SQL Server}~\cite{larson2015real} aims to ensure data consistency during synchronization and avoid write conflicts with user transactions. This technique uses incremental storage and a delete table for synchronization, divided into two phases:
(1) Scan the incremental storage for data not yet synchronized to the columnar index. Frequently updated hot data (typically within the last hour) is temporarily excluded to avoid performance degradation. Selected data tuples are encoded into columnar format and stored in the columnar table, while each row is assigned a Row ID (RID) and inserted into the "Deleted Rows Table" to logically hide these rows. After this phase, data is physically migrated to the columnar index but remains logically invisible.
(2) A series of background transactions delete the data in the delete table. Once these transactions complete, the tail data in the incremental storage is cleared, and the new data in the columnar index becomes visible to subsequent queries. To avoid write-write conflicts with user transactions, SQL Server employs the following strategies: If the background transaction has not committed, it is treated as the victim in case of conflict (i.e., the background transaction is aborted). If the background transaction has committed, user transactions are allowed to overwrite the changes made by the background transaction, as these changes do not affect user-visible data.

Two-phase migration effectively addresses the synchronization of incremental data with columnar data by separating data migration and update visibility.

\textbf{Threshold-Driven Synchronization}~\cite{lahiri2015oracle} determines the timing of synchronization from incremental storage to baseline columnar data. This technique uses a transaction journal to record the Row IDs of all changed data. When the number of changes reaches a set threshold, the system retrieves the latest data from the incremental row storage based on the Row IDs and merges it into the columnar storage. However, if the threshold is too large, the freshness of the columnar data may decrease. To address this, this approach periodically merges data from the incremental storage into the columnar storage to ensure consistency and real-time performance.

\textbf{Dictionary-Sorted Merge in SAP HANA}~\cite{sikka2012efficient}
allows direct synchronization of two dictionary-encoded datasets, avoiding the overhead of decoding and re-encoding. This technique consists of two steps:
(1) Append new row data to the incremental columnar storage. For each new column, the system queries the dictionary in the incremental columnar storage. If the data exists in the dictionary, its corresponding encoded value is added to the data column. If not, the dictionary is updated, a new encoding is assigned, and the new data is added to the data column.
(2) Merge the dictionary in the incremental columnar storage with the full dictionary in the columnar storage. Through insertion sorting, a merged full dictionary is obtained. The data in the incremental columnar storage is then added to the full data using the new dictionary encoding, completing the synchronization.

\section{Conclusion}

This work presents a novel storage engine, \xstore, designed to address the inefficiency of update operations in columnar storage structures under hybrid workload scenarios. While columnar storage demonstrates excellent query performance when handling large-scale datasets, its update operations often suffer from complexity and poor real-time performance in data-intensive application scenarios. \xstore innovatively integrates an incremental row storage mechanism with a fine-grained row-to-column transformation and compaction strategy, successfully optimizing the balance between data update efficiency and query performance. Specifically, the design architecture of \xstore includes the following key features: (1) A collaborative mechanism between incremental row storage and columnar storage; (2) A fine-grained row-to-column transformation and compaction strategy; (3) A cost-based intelligent scheduling mechanism; (4) Multi-version read and concurrency control mechanisms.

Experimental evaluation results show that, compared to existing columnar storage systems (such as DuckDB), \xstore not only significantly improves update performance under hybrid workloads but also achieves considerable improvements in query performance.

\bibliographystyle{splncs04}
\bibliography{reference}

\end{document}